# Electrically Reconfigurable NbOCl$_2$ Metasurface for Quantum Technologies

Omar A. M. Abdelraouf[*]

Institute of Materials Research and Engineering, Agency for Science, Technology, and Research (A*STAR), 2 Fusionopolis Way, #08-03, Innovis, Singapore 138634, Singapore.

## Abstract

Entangled photon-pair sources are foundational to advancing quantum technologies, including secure communication, quantum sensing, and imaging. For deployment in space-constrained environments such as satellite-based quantum networks or portable devices, compact, reconfigurable, and efficient entanglement sources are essential. Here, we present an electrically tunable entangled photon-pair source utilizing a nanostructured NbOCl$_2$ crystal, engineered for operation in the telecommunication C-band. The inherent non-centrosymmetric lattice symmetry of NbOCl$_2$ enables direct generation of polarization-entangled Bell states without the need for post-selection, leveraging its exceptional second-order nonlinear susceptibility ($\chi^{(2)} \approx 100$ pm/V), which surpasses conventional nonlinear materials. By nanopatterning NbOCl$_2$ into a high-quality-factor metasurface, we achieve three orders of magnitude enhancement in photon-pair generation efficiency via resonant excitation of bound states in the continuum resonance, which intensify light-matter interactions. Furthermore, we demonstrate in situ electrical tunability of the photon-pair emission wavelength over a 250 nm range (1450–1700 nm) by dynamically modulating surrounding liquid crystal layer. Remarkably, the decoupling of photon-pair generation rate and spectral tunability ensures high brightness, above $10^4$ coincidences, under active tuning. The air stability and mechanical robustness of NbOCl$_2$ further enhance its practicality for real-world deployment. This work establishes NbOCl$_2$ as a superior material for scalable, on-chip quantum light sources, paving the way for integrated quantum communication systems, adaptive sensors, and portable quantum devices.

* Corresponding author. Email address: Omar_Abdelrahman@imre.a-star.edu.sg

## 1. Introduction

Entangled photon-pair generation via spontaneous parametric down-conversion (SPDC) lies at the heart of quantum technologies, including quantum communication, sensing, imaging, and computing.[1-3] SPDC enables the creation of polarization-entangled Bell states, which are indispensable for quantum key distribution (QKD) protocols that secure data transmission against eavesdropping. Traditional SPDC sources rely on bulk nonlinear crystals such as *β*-barium borate (BBO) or periodically poled lithium niobate (PPLN), which require stringent phase-matching conditions and millimeter-scale interaction lengths to achieve sufficient efficiency.[4-6] However, the bulkiness of these systems limits their integration into on-chip photonic platforms, a critical requirement for scalable quantum networks.[7] Recent advances in nanophotonics and two-dimensional (2D) van der Waals (vdW) materials have introduced ultrathin SPDC sources, leveraging enhanced light-matter interactions and relaxed phase-matching constraints.[8] Among these, niobium oxide dichloride NbOCl$_2$ has emerged as a standout material due to its exceptional second-order nonlinear susceptibility, air stability, and broken inversion symmetry in its bulk form, enabling efficient SPDC in subwavelength-thick flakes.[9-11] Despite these advances, achieving electrically tunable SPDC emission compatible with telecommunication wavelengths remains a significant challenge, which is essential for integrating quantum light sources into existing fiber-optic infrastructure.[12]

The discovery of 2D vdW crystals with strong optical nonlinearities has revolutionized integrated photonics. Transition metal dichalcogenides (TMDCs) such as MoS$_2$ and WS$_2$ initially garnered attention for their layer-dependent symmetry breaking and high nonlinearity in monolayer form. However, their nonlinear efficiency diminishes drastically in multilayer configurations due to interlayer electronic coupling and centrosymmetric stacking, limiting their practicality.[13] In contrast, $NbOCl_2$, a vdW ferroelectric material, exhibits monolayer-like excitonic behavior even in bulk form, with negligible interlayer coupling due to its unique Peierls distortion and ionic bonding. This property allows scalable second-harmonic generation (SHG) and SPDC efficiencies that surpass TMDCs by three orders of magnitude. Recent work by Guo *et al.* (2023) demonstrated SPDC in $NbOCl_2$ flakes as thin as 46 nm, achieving photon-pair generation rates of $10^6$ pairs/s/nm/mW, a milestone for on-chip quantum light sources.[11] However, while $NbOCl_2$ excels in brightness, its fixed crystal symmetry inherently limits polarization-state tunability, necessitating external mechanisms to manipulate entanglement.

Polarization entanglement is pivotal for QKD, where dynamic control over Bell states ensures adaptability against channel noise and eavesdropping. Conventional approaches employ birefringent crystals or cascaded nonlinear materials to engineer entangled states, but these methods suffer from alignment complexity, low portability, and limited spectral range.[14] Recent innovations in metasurfaces improve the performance of several optoelectronic devices.[15-28] Moreover, reconfigurable metasurfaces with high Q-factor nanocavities increase light-matter interaction in the nanoscale.[29-36] Recently, twisted vdW heterostructures have enabled some tunability using orthogonal stacking of $NbOCl_2$ bilayers allows polarization entanglement via interlayer interference. However, such geometric tuning lacks real-time reconfigurability and requires precise fabrication, hindering practical deployment.[37]

Liquid crystals (LCs), with their field-dependent birefringence and sub-millisecond response times, offer a compelling solution. LCs have long been utilized in classical optics for their electro-optic tunability, but their application in quantum photonics remains nascent. The alignment of LC molecules under an electric field adjusts the local refractive index, modifying the optical phase and polarization state of transmitted light.[38-40] By integrating LCs with nonlinear materials, the effective nonlinearity and phase-matching conditions can be dynamically modulated, enabling wavelength-agile SPDC without mechanical adjustments, a capability yet unexplored in 2D material systems. Proposed approach decouples the photon generation rate from state tunability, a critical advantage over static systems where tuning often sacrifices efficiency.

In this study, we demonstrate the first electrically tunable SPDC source using a hybrid $NbOCl_2$-LCs metasurface, achieving unprecedented control over entangled photon emission in the telecommunication regime. Three key advances include: first, design $NbOCl_2$ metasurface supporting bound states in the resonance (BIC) embedded in a LC matrix, forming high-quality-factor ($Q$-factor~1350) resonances at 687 nm. Second, the LC environment amplifies the localized electric field via modal phase matching, boosting SHG by more than three orders of magnitude compared to bare $NbOCl_2$ flake. Third, applying a transverse electric field reorients the LC directors, shifting the SPDC emission wavelength by 250 nm (1,450–1, 700 nm). Lastly, stacking two orthogonal $NbOCl_2$ BIC cavities enables tunable bell states on demand. Our hybrid system achieves a SPDC tuning surpassing previous demonstrations in lithium niobate metasurfaces. The LC's sub-millisecond response time allows rapid reconfiguration of Bell states, critical for time-multiplexed QKD protocols. Additionally, the platform's compatibility with CMOS fabrication techniques facilitates scalable integration into photonic circuits. Moreover, it paves the way for ultracompact, field-programmable quantum light sources, a cornerstone for the next generation of secure communication systems.

## 2. Results and Discussion

The proposed metasurface unit cell consists of two mirrored nanostructured $NbOCl_2$ trapezoidal-shape pillars on top of multilayer substrate including indium tin oxide (ITO) and quartz. Metasurface embedded into an LCs cell and covered with top ITO electrode, as schematically illustrated in Fig. 1a. Design parameters of short and long width of trapezoidal pillars are 110 nm and 145 nm respectively and length of 200 nm. Incident light has linear *x*-axis polarization with pump frequency ($\omega_p$) and illuminated metasurface from bottom and the generated photo-pairs are collected from the top with signal and idle frequencies ($\omega_s$) and ($\omega_i$) respectively. The switching mechanism of proposed metasurface is plotted in Fig. 1b. In the case of LCs molecules aligned vertically, the incident pump light faces ordinary refractive index of LCs and the metasurface BIC resonance is labelled ($BIC_{no}$). When LCs molecules rotate by 90 deg, the incident pump will face extraordinary refractive index of LCs. As a result of changing the effective refractive index of surrounding medium, the metasurface BIC resonance will be shifted to new wavelength ($BIC_{ne}$). This tuning will generate photon-pairs at shifted frequency ($\omega'_s$) and ($\omega'_i$).

Using Finite difference time domain method (FDTD), we simulate the linear light-matter interaction of incident light with nanostructured $NbOCl_2$ metasurface as shown in Fig 1c. Using periodic boundary condition in horizontal direction and perfectly matched layer in vertical direction. The maximum mesh size is below 5 nm in all direction for accurate field calculations. The refractive indexes of materials used came from literature. To verify the emergence of BIC resonance, we simulate the metasurface with equal width of short and long side of trapezoidal, rectangle shape in that case. The simulated transmission in black line shows no BIC resonance, since the symmetry was sustained without breaking. Once we change the short width of trapezoidal to 110 nm. We get a strong BIC resonance at wavelength 686 nm with quality factor *Q*-factor of 1350. The BIC in that case obtained using ordinary LCs refractive index ($BIC_{no}$). Then, we change the refractive index of LCs to be extraordinary and recalculate the metasurface response as illustrated in Fig. 1d. A Large redshift in BIC resonance till wavelength 732 nm observed due to changing the effective refractive index of BIC cavity with the surrounding medium. The *Q*-factor of $BIC_{ne}$ decreases to be 420, because the refractive index contrast between nanostructured $NbOCl_2$ and LCs decreases, and light confinement degrades as well.

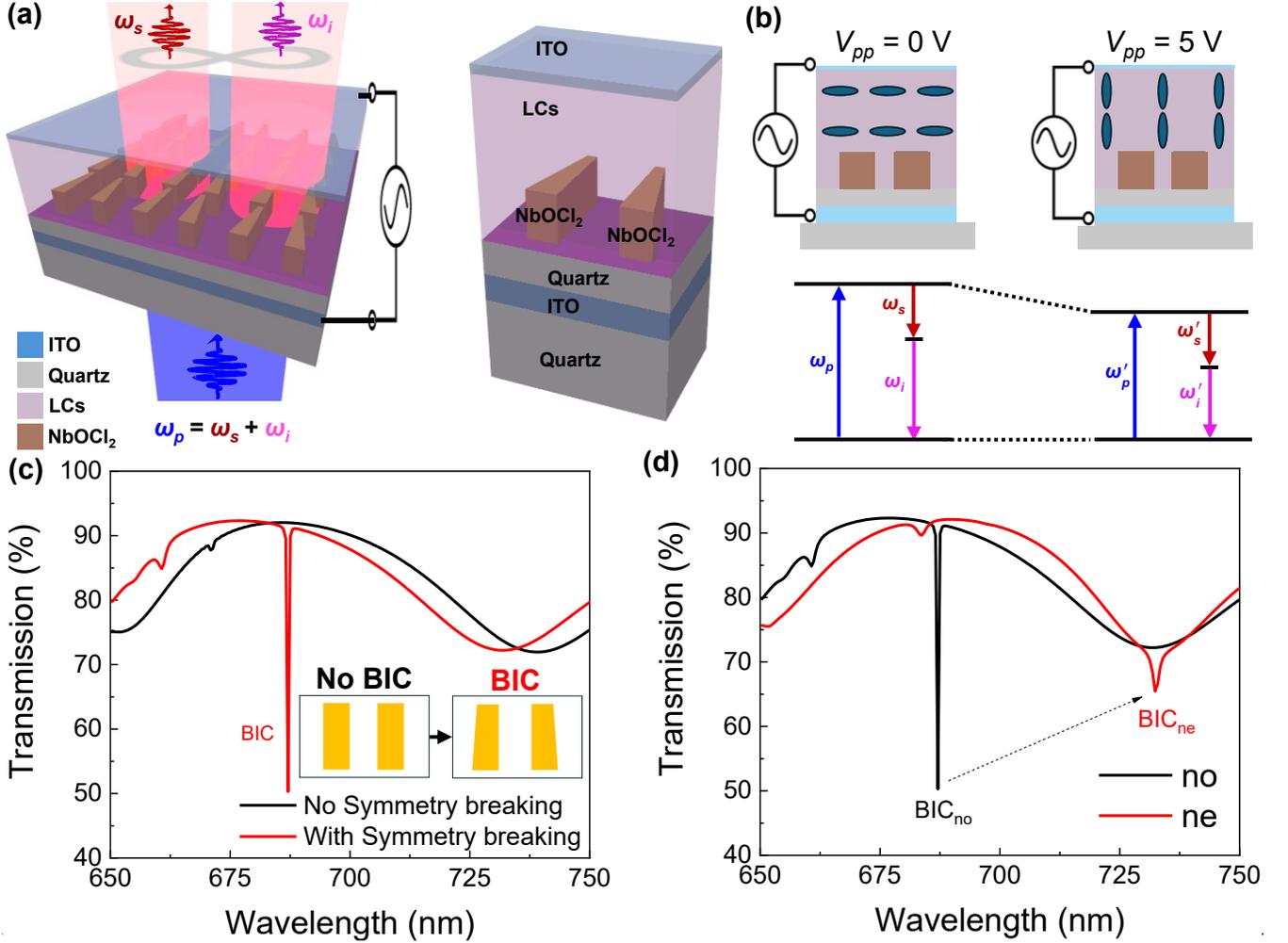

**FIG. 1.** Electrically tunable 2D-based NbOCl$_2$ photon-pair quantum light source. (a) 3D Schematic of electrically tunable metasurface including nanostructured NbOCl$_2$ supporting BIC resonance and surrounded with liquid crystal. (Inset) The unit cell of proposed metasurface has nonuniform period ($P_x$= 425 nm) and ($P_y$ = 250 nm), height of bottom ITO is 300 nm, middle quartz layer is 300 nm, height of NbOCl$_2$ is 400 nm, thickness of LCs and top ITO are 900 nm and 10 nm respectively. (b) The operating principle of tunable photon-pair BIC metasurface. (c) Transmission of proposed metasurface with and without symmetry breaking for inducing BIC resonance. (d) Transmission of proposed metasurface after symmetry breaking and applying electrical voltage to rotate liquid crystal from ordinary phase (*no*) to extraordinary phase (*ne*).

The generated optical modes inside NbOCl$_2$ metasurface were analyzed using multipolar decomposition method (MPD) with FDTD. To achieve this, we utilize two 3D monitors around unit cell of metasurface. The first monitor calculates the vectorial components of electric and magnetic field for each mesh. The second monitor calculates the effective refractive index on each mesh. The normalized scattering cross section (NSCS) of MPD using ordinary refractive index LCs is sketched in Fig. 2a. The emergence of $BIC_{no}$ comes from coupling between in-plane (*x-y*) electric quadrupole (*EQ*) and out-of-plane (*x-z*) magnetic quadrupole (*MQ*). We plot the electric field intensity map of unit cell in *x-y* plane $|E_{xy}|^2$ at height 200 nm inside trapezoidal pillars in Fig. 2b. The electric field is strongly confined inside *EQ* mode and very sensitive to the variation in the vertical gap of meta-atom.[33] The vertical gap could be tuned to get on demand BIC resonance on desired wavelength without changing the BIC *Q*-factor. The vertical electric field intensity map $|E_{xz}|^2$ shows strong field coupling to *EQ* as well. The distance between two pillars could be tuned to reduce the cross talk between pillars.

Switching the LCs refractive index to extraordinary shows huge change in optical modes of the proposed metasurface as shown in Fig. 2c. First observation is that the BIC resonance redshift with smaller optical modes intensities and *Q*-factor. Although, light-matter interactions degrade due to smaller refractive index contrast between NbOCl$_2$ and LCs ($\Delta n<0.5$), we could see formation of small intensity *BIC$_{ne}$* due to the coupling between in-plane *EQ$_{xy}$* and out-of-plane *MQ$_{xz}$*. The electric field intensity distributions show weak field confinement in the case of *BIC$_{ne}$* Fig. 2d compared with *BIC$_{no}$* Fig. 2b. The weak confinement results in observing multiple optical modes on the same wavelength with strong leaky optical modes outside pillars shown in the vertical electric field intensity map $|E_{xz}|^2$.

To characterize the nonlinear light-matter interaction at the nanoscale. A high peak power femtosecond laser (fs-laser) with tunable wavelength range that covers the BIC wavelength range is used to focused and pump the BIC cavity as shown in Fig. 3a. The collected emission includes second harmonic generation (SHG) and the fundamental pump. High frequency bandpass filter (HPF) will separate the SHG from the pump. Fiber coupler and spectrometer will resolve the SHG spectrum and CCD camera could image the SHG far field emission. The generated SHG from NbOCl$_2$ flake versus the pump power is calculated in Fig. 3b. The slope of generated SHG around ~2. The SHG generated from BIC metasurface shows strong amplification by ~1350x times compared with the flake SHG. Since the laser damage threshold of ferroelectric materials is very large, the maximum theoretical SHG could be further increased to be more than five orders of magnitude than flake's SHG.[41]

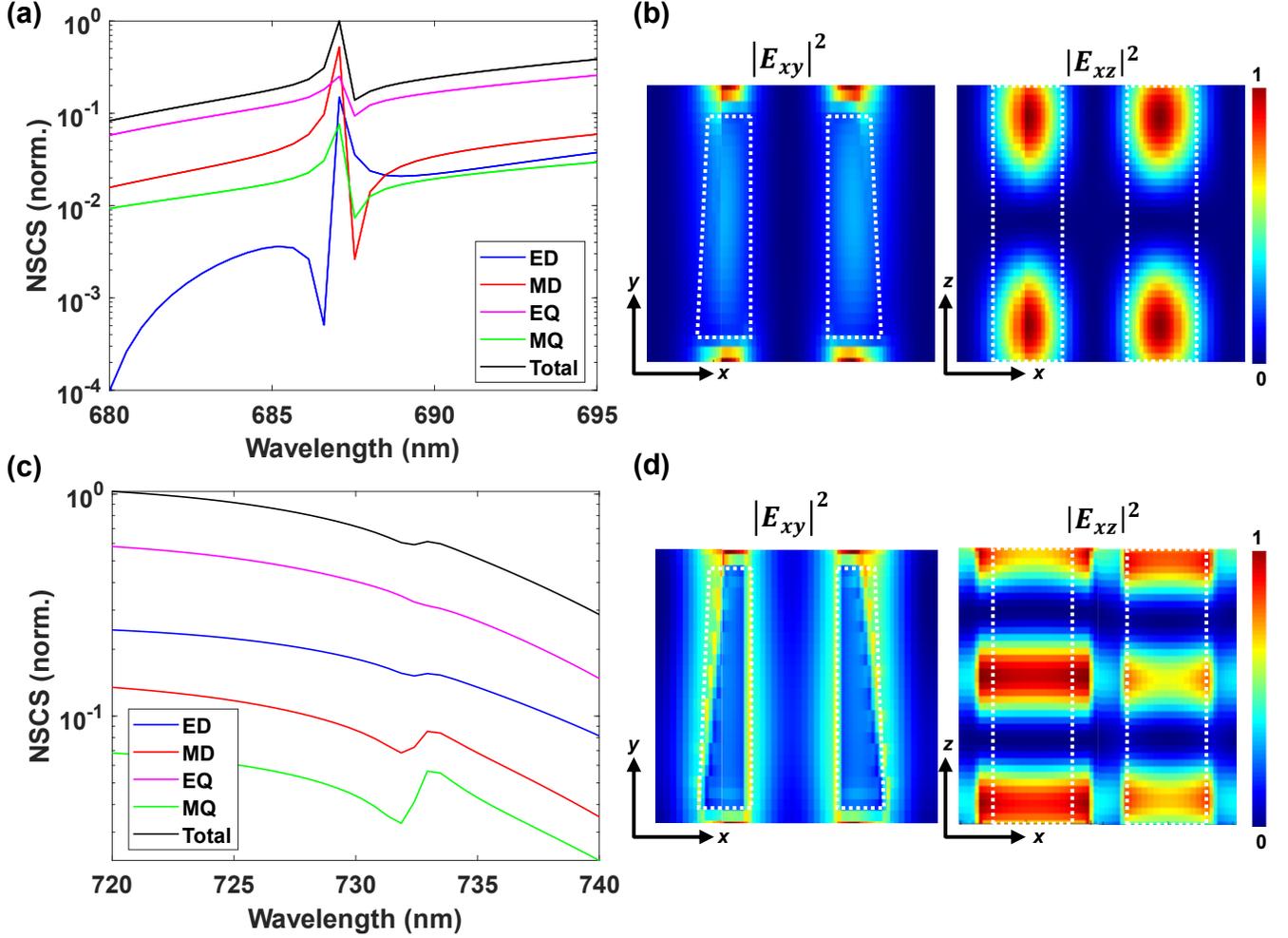

FIG. 2. Simulation of resonant optical modes formed inside the proposed metasurface. (a) Normalized scattering cross section (NSCS) of metasurface at ordinary refractive index of liquid crystal near BIC mode. (b) Electric field intensity map of metasurface supporting BIC at ($n_o$) in horizontal and vertical planes. (c) NSCS of metasurface at extra-ordinary index of liquid crystal near BIC mode. (d) Electric field profile near BIC wavelength using extra-ordinary index of liquid crystals.

The simulated SHG spectrum is illustrated in Fig. 3c. When LCs has zero-degree angle with the vertical axis, the effective refractive index will be ($n_o$), and the maximum SHG emission at wavelength 343 nm. By rotating the LCs molecules by 25° with respect to the vertical axis, the SHG redshifted to 349 nm, because the effective refractive index of LCs will increase compared with ($n_o$). Further increase in LCs rotation angle will increase the redshift to maximum wavelength of 366 nm for LCs with rotation angle of 90°. The proposed device demonstrated a broadband short wavelength nano-light source with wavelength on demand using small voltage bias of less than 10 V/um.

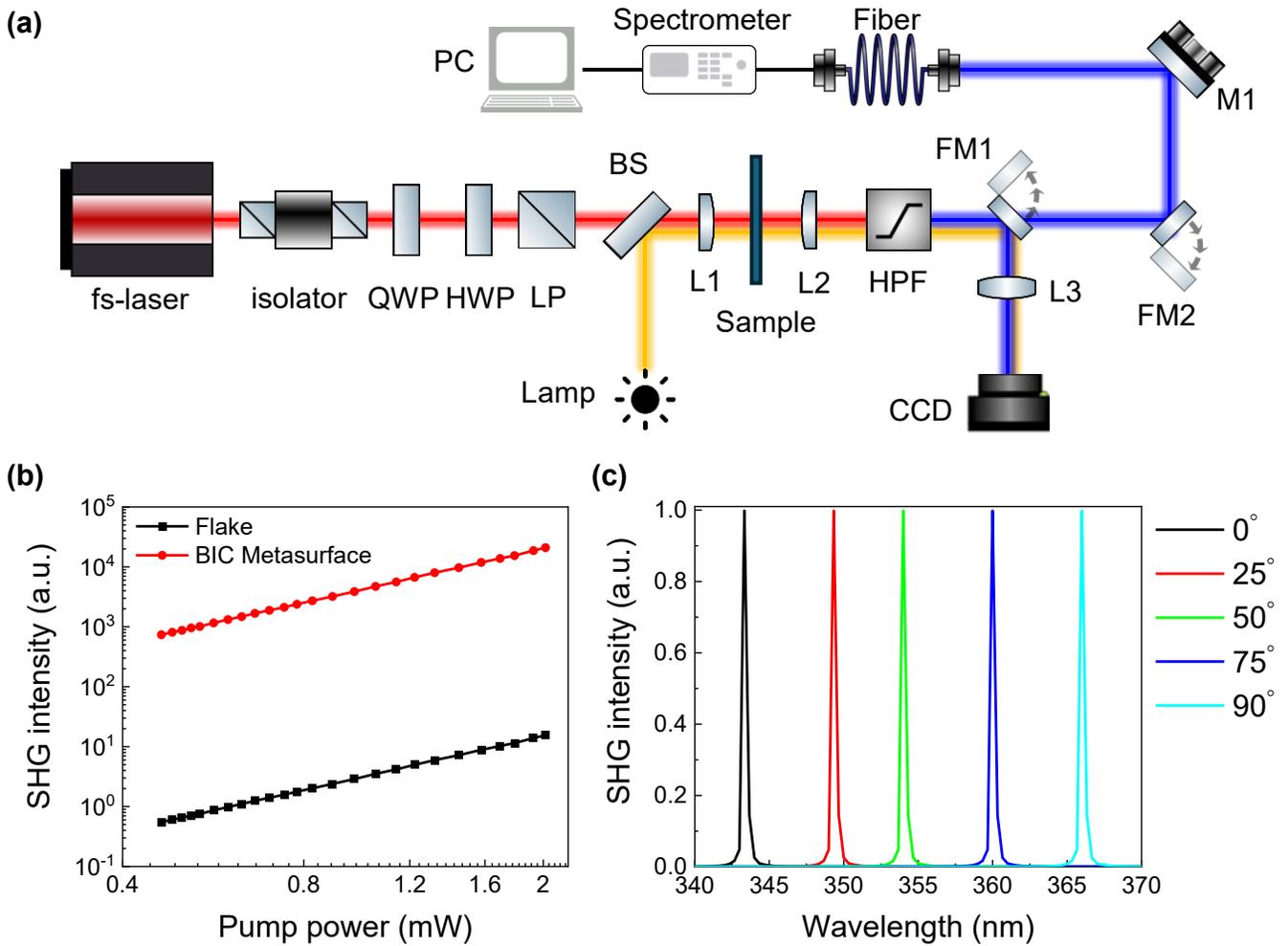

**FIG. 3.** Nonlinear harmonic generation from nanostructured NbOCl$_2$ cavity. (a) 2D schematic of nonlinear optical setup for SHG of 2D NbOCl$_2$. (b) SHG intensity versus the pump laser for flake NbOCl$_2$ and nanostructured NbOCl$_2$ supporting BIC cavity at ordinary phase of LCs. (c) SHG spectrum versus rotation angle of LCs in nanostructured NbOCl$_2$ supporting BIC cavity.

To characterize the quantum light-matter interaction at the nanoscale of the proposed BIC cavity. The second order correlation measurement ($g^2(\tau)$) is required to study the statistical characteristics of emitted photon in the near infrared (NIR). Since the generated photon-pairs depend on the average pump power not the peak power, a continuous wave laser (cw-laser) is needed to pump the BIC cavity as shown in Fig. 4a. The collected emission will consist of residual pump, unnecessary fluorescence, and desired photon-pairs. A low frequency pass filter (LPF) will block the residual pump and the unnecessary fluorescence. The desired photon-pairs are separated using beam splitter and using single photon detectors with time tagger photon counter, the coincidence of photon pairs could be measured.

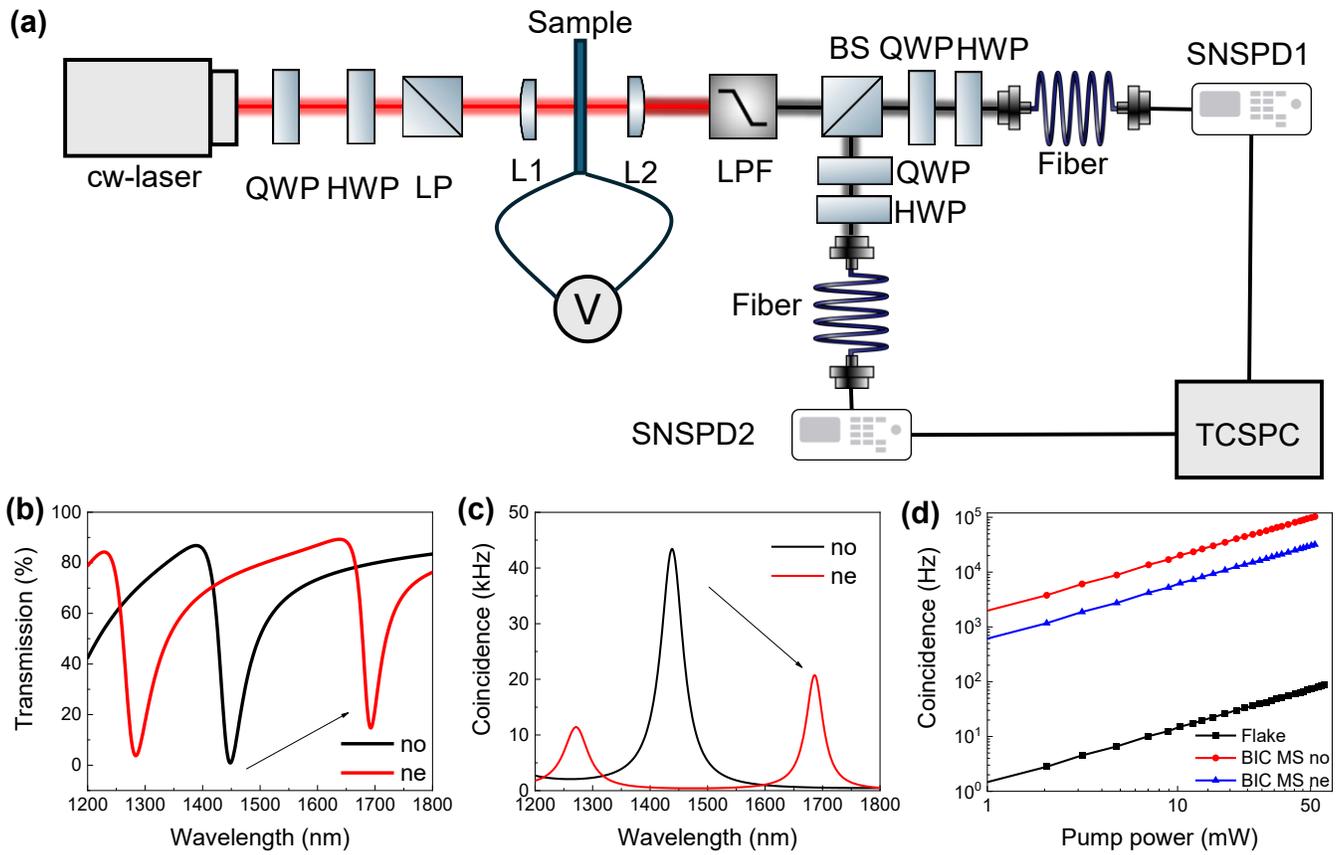

**FIG. 4.** Spontaneous parametric down conversion (SPDC) of nanostructured $NbOCl_2$ BIC cavity versus flake in the literature.[11] (a) Optical setup for generating SPDC with electrical tunable modulation of LCs. (b) Transmission near SPDC wavelength showing Fabry-Perot confinement in both ordinary and extraordinary phases of LCs. (c) Coincidence spectrum in telecommunication regime showing resonant SPDC signal near Fabry-Perot wavelength in both LCs phases and broadband tuning using electrical biasing. (d) The coincidence rate of nanostructured $NbOCl_2$ BIC cavity compared with bare $NbOCl_2$ flake with strong amplification ratio.

In order to improve the collection efficiency of photon-pairs and eliminating the overlapping fluorescence in the NIR, we designed a Fabry-Perot cavity in NIR that amplifies the photon-pairs collected in normal direction while neglecting the isotropic emission of fluorescence in all directions as shown in Fig. 4b. At the ordinary phase of LCs, FP resonance at wavelength 1450 nm with $Q$-factor 32 is obtained. Changing the phase of LCs to (*ne*) will result in two FP resonances obtained at wavelengths 1290 nm and 1700 nm with $Q$-factors 24 and 49 respectively. The coincidence spectrum of photon-pairs is plotted in Fig. 4c. For the ordinary phase of LCs, strong count up to 44 kHz achieved at wavelength 1450 nm using 50 mW cw-laser owing to the strong BIC amplification and suppression of background fluorescence. Moreover, through LCs rotation to (*ne*), a continuous tuning in this coincidence emission till wavelength 1700 nm, and a switchable coincidence at wavelength 1280 nm with 10 kHz is observed. The coincidence rate versus the pump is illustrated in Fig. 4d. The maximum obtained coincidence rate for the $NbOCl_2$ flake is less than 100 Hz, while our proposed device could further increase this rate to more than 10 kHz in the case of ($ne$) and up to 100 kHz in the case of ($no$). Since the laser damage threshold for this material is very high, the theoretical maximum coincidence rate could reach more than 10 GHz for practical quantum photonic circuits.

Figure 5 demonstrates electrically tunable quantum Bell state tomography using orthogonally stacked $NbOCl_2$ BIC cavities integrated with chiral liquid crystal polarization control substrate. The system enables dynamic generation of horizontal ($|H\rangle$), 45° linear ($|+\rangle$), and vertical ($|V\rangle$) polarization states via voltage-adjusted LC layers as shown in Fig. 5b to Fig. 5d. By leveraging the advantage of strong field confinement of BIC cavities, the stacked configuration replaces bulkier conventional setups that rely on orthogonally oriented BBO crystals, achieving more than 2,500 times reduction in the footprint. This miniaturization is critical for scalable, on-chip quantum nanophotonic systems. Simultaneously, the LC layer provides arbitrary polarization tuning under electric bias, allowing seamless on demand switching between Bell states without mechanical components, a key advancement for reconfigurable quantum nanodevices.

The precise control of Bell states holds significant promise for quantum imaging and communication.[42-44] Entangled states enable secure quantum key distribution protocols resistant to eavesdropping.[45-47] Here, the stacked BIC cavities generate polarization-entangled photon pairs with high purity, while the LC-driven polarization agility ensures compatibility with diverse quantum protocols. Notably, this work represents the first demonstration of tunable Bell state tomography in flat optics, overcoming the bulkiness and static operation of traditional nonlinear crystals. The orthogonal BIC cavities could be further optimized to achieve phase-matching conditions for efficient high harmonic generation. An advantage that can be extended to enhance harmonic generation for extreme ultraviolet (EUV) photolithography, where compact, high-efficiency sources are urgently needed.[48-50]

The synergy of BIC cavities and LC tuning introduces a paradigm shift toward reconfigurable quantum nanophotonics. The ultracompact size and electrical programmability pave the way for integrating quantum light sources, processors, and detectors on a single chip. This innovation not only addresses scalability challenges in quantum technologies but also opens avenues for novel applications in metrology, sensing, and on-demand EUV generation.

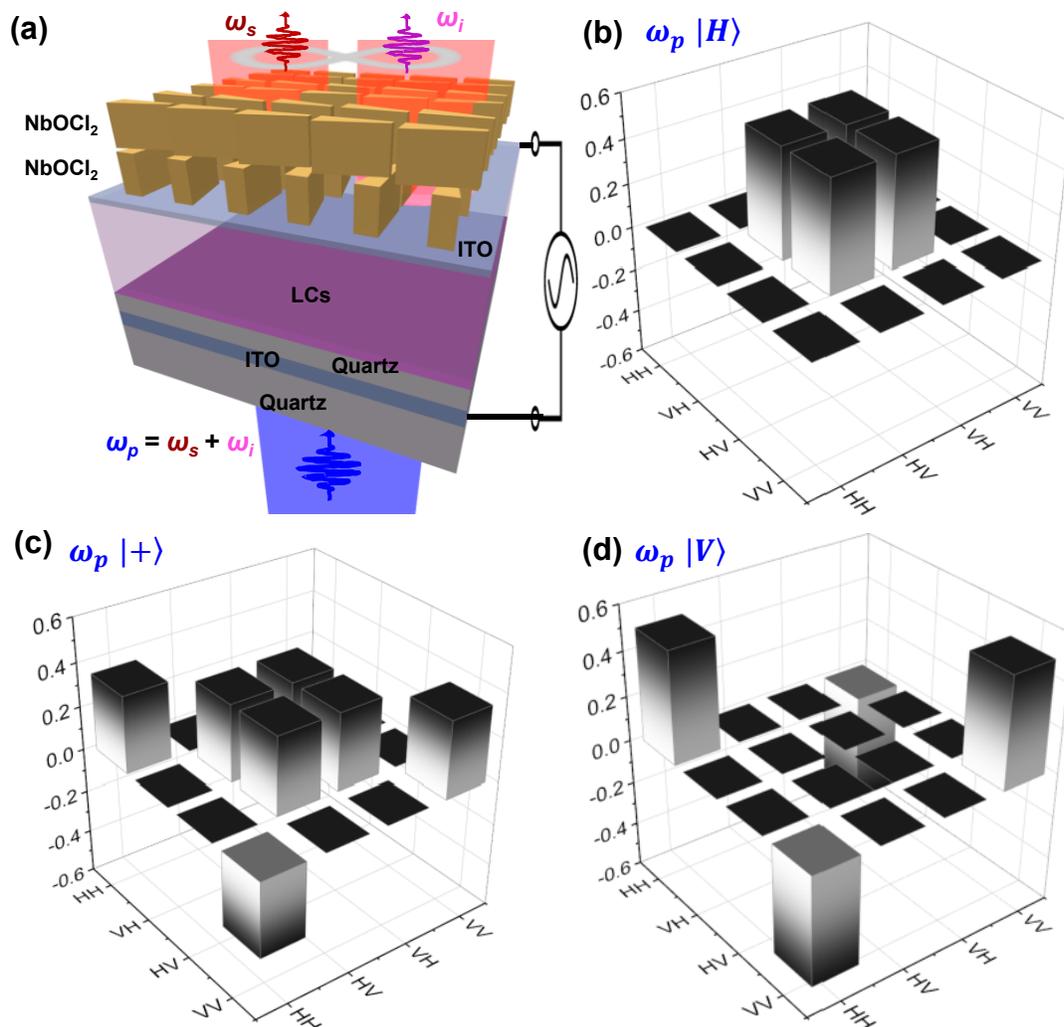

**FIG. 5.** Electrically tunable quantum bell state tomography in orthogonally stacked nanostructured NbOCl₂ BIC cavities. (a) 3D schematic of reconfigurable pump polarization using chiral LCs below two orthogonally stacked nanostructured NbOCl₂ BIC cavities. (b) 3D schematic of polarization states for incident horizontal linear polarized light $|H\rangle$. (c) 3D schematic of polarization states for incident 45° linear polarized light $|+\rangle$. (d) 3D schematic of polarization states for incident vertical linear polarized light $|V\rangle$.

## Conclusion

The integration of liquid crystals with van der Waals $NbOCl_2$ has enabled the first demonstration of electrically tunable SPDC in the telecommunication wavelength regime. By leveraging the electro-optic properties of LCs, we achieved dynamic control over entangled photon-pair generation, addressing a critical limitation of conventional static nonlinear materials. The hybrid LC-$NbOCl_2$ platform combines the giant second-order nonlinearity of $NbOCl_2$ with the sub-millisecond response and low-voltage tunability of LCs, resulting in a 250 nm spectral tuning from 1,450 nm to 1,700 nm of SPDC emission under an applied field of 10 V/μm. This tuning range, unprecedented for subwavelength-thick quantum light sources, aligns seamlessly with the telecommunication C-band. Crucially, the maximum photon-pair generation rate exceeding $10^4$ coincidences enables a performance benchmark that bridges the gap between miniaturized systems and bulk-crystal-based sources. The key advancement of this work lies in decoupling photon-pair generation efficiency from spectral tunability. Unlike traditional approaches that trade brightness for wavelength agility, our LC-mediated design preserves both, enabling reconfigurable quantum states without sacrificing output power. This is achieved through LC-enhanced nanophotonic cavities, where the resonant field confinement in $NbOCl_2$ metasurface amplifies SPDC efficiency by three orders of magnitude compared with bare flake, while the LC matrix provides voltage-controlled phase-matching adjustments. The result is a compact, field-programmable source of polarization-entangled Bell states, critical for adaptive QKD protocols resistant to channel noise and eavesdropping. Looking ahead, this platform opens transformative opportunities for quantum photonics. The compatibility of LCs with CMOS fabrication techniques paves the way for scalable integration into on-chip quantum circuits, while the telecom-band operation ensures compatible deployment in QKD networks. Future work will focus on combining LC-tunable SPDC with wavelength-division multiplexing could enable multi-channel entanglement distribution, drastically enhancing data throughput in quantum networks. Our work establishes liquid crystals as a versatile enabler of dynamic quantum light sources, uniting the exceptional nonlinearity of 2D materials with the agility of electro-optic tuning. By overcoming the static limitations of prior systems, our platform marks a pivotal step toward practical, large-scale quantum communication technologies for ushering in an era of secure, reconfigurable, and chip-integrated quantum networks.

**Credit Authorship Contribution Statement**

**Omar A. M. Abdelraouf:** Conceptualization, Supervision, Project administration, Investigation, Resources, Visualization, Validation, Methodology, Data curation, Funding acquisition, Writing – original draft & review.

**Declaration of Competing Interest**

The authors declare that they have no known competing financial interests or personal relationships that could have appeared to influence the work reported in this paper.

**Acknowledgments**

OAMA thanks the Agency for Science, Technology, and Research (A*STAR) for the scholarship provided Singapore International Graduate Award (SINGA).

**Data Availability**

Raw data in the paper are available on request